\documentclass[11pt,letterpaper]{article} % Remove this line if using A4 size.
\usepackage[
  letterpaper,
  left=0.6in,right=0.6in,top=0.6in,bottom=0.6in,
  headheight=85pt,headsep=-45pt
]{geometry}
\usepackage{graphicx}
\graphicspath{{./}{figures/}}
\usepackage{float}
\usepackage{amsmath}
\usepackage{tikz}

\usepackage{caption}
\usepackage{booktabs}
\usepackage{array}
\usepackage{tabularx}
\usepackage{enumitem}
\setlist{nosep}
\usepackage[table]{xcolor}
\definecolor{tableheader}{HTML}{D9D9D9}
\newcolumntype{P}[1]{>{\sffamily\centering\arraybackslash}p{#1}}
\newcolumntype{Y}{>{\sffamily\centering\arraybackslash}X}
\newcolumntype{A}[1]{>{\raggedright\arraybackslash}p{#1}}
\newcolumntype{M}[1]{>{\raggedright\arraybackslash}m{#1}}

\usepackage[T1]{fontenc}
\usepackage[utf8]{inputenc}
\usepackage{PTSerif}
\usepackage[scaled]{helvet}

\newcommand{\headingfont}{\sffamily}

\usepackage{titlesec}
\titleformat{\section}
  {\headingfont\bfseries\raggedright\fontsize{18pt}{18pt}\selectfont}{}{0.75em}{}
\titleformat{\subsection}
  {\headingfont\bfseries\raggedright\fontsize{15pt}{16pt}\selectfont}{}{0.75em}{}
\titleformat{\subsubsection}
  {\headingfont\bfseries\raggedright\fontsize{12pt}{14pt}\selectfont}{}{0.75em}{}

\titlespacing*{\section}{0pt}{7pt}{6pt}
\titlespacing*{\subsection}{0pt}{7pt}{6pt}
\titlespacing*{\subsubsection}{0pt}{7pt}{6pt}

\newcommand{\miniheading}[1]{%
  \par\noindent{\headingfont\bfseries\fontsize{12pt}{14pt}\selectfont #1}\par\vspace{4pt}%
}

\usepackage{changepage}

\newenvironment{narrow}
{\begin{adjustwidth}{1cm}{1cm}}
{\end{adjustwidth}}

\usepackage{fancyhdr}
\fancyhfoffset[R]{18pt}
\fancypagestyle{firstpage}{
  \fancyhf{}
  \fancyhfoffset[R]{18pt}
  \fancyhead[R]{%
    \smash{\raisebox{0pt}[0pt][0pt]{
      \begingroup\setlength{\fboxsep}{20pt}
        \colorbox{white}{\includegraphics[height=0.6in]{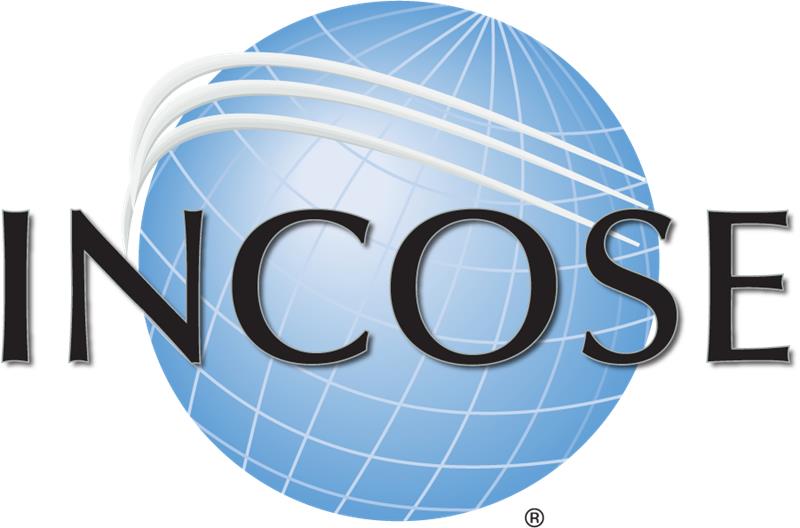}}%
      \endgroup
    }}
  }
  \fancyfoot[R]{\sffamily\bfseries\footnotesize \thepage}

}
\makeatletter
\@ifundefined{theauthor}{}{}
\makeatother
\usepackage{titling}
\pretitle{\headingfont\bfseries\fontsize{24pt}{26pt}\selectfont\raggedright}
\posttitle{\par\vspace{-.3in}}
\preauthor{}\postauthor{}
\author{\mbox{}}
\date{}
\newcommand{\authorcard}[5]{%
  {\headingfont\bfseries\fontsize{12pt}{14pt}\selectfont #1}\par
  {\headingfont\bfseries\fontsize{12pt}{14pt}\selectfont #2}\par
  {\headingfont\bfseries\fontsize{12pt}{14pt}\selectfont #3}\par
  {\headingfont\bfseries\fontsize{12pt}{14pt}\selectfont #4}\par
  {\headingfont\bfseries\fontsize{12pt}{14pt}\selectfont #5}\par
}



\makeatletter
\newcommand{\colfig}[2][]{%
  \IfFileExists{#2}{\includegraphics[width=\linewidth,#1]{#2}}{%
    \fbox{\parbox[b][1.5in][c]{\linewidth}{\centering \textit{Missing figure: }#2}}}%
}
\makeatother

\usepackage{csquotes}
\usepackage[style=apa,backend=biber]{biblatex}
\usepackage{changepage}

\usepackage{multicol}
\usepackage[hidelinks]{hyperref}

\title{Towards Formalising Stakeholder\\ Context using SysML v2}

\begin{document}
\maketitle
\thispagestyle{firstpage}

% ---- Authors ----
% ---- For the initial paper submission, do not include any author information. For the final paper submission, format author information as shown below. ------------------

\noindent
\begin{tabular*}{\textwidth}{@{\extracolsep{\fill}} A{0.32\textwidth} A{0.32\textwidth} A{0.32\textwidth}}
  \authorcard{Matthew Harrison}{Loughborough University}{Epinal Way} {Loughborough, Leicestershire, LE11 3TU}{M.Harrison2@lboro.ac.uk} &
  \authorcard{John Carlin}{Nuclear Waste Services}{Pelham House, Pelham Drive} {Calderbridge, Cumbria, CA20 1DB}{john.carlin@\newline nuclearwasteservices.uk} &
  \authorcard{Chengyuan Liu}{Loughborough University}{Epinal Way} {Loughborough, Leicestershire, LE11 3TU}{C.Liu10@lboro.ac.uk} \\
  \multicolumn{3}{@{}c@{}}{\rule{0pt}{0.9\baselineskip}} \\[-0.2\baselineskip]
  \authorcard{Sarah Dunnett}{Loughborough University}{Epinal Way} {Loughborough, Leicestershire, LE11 3TU}{S.J.Dunnett@lboro.ac.uk} &
  \authorcard{Siyuan Ji}{Loughborough University}{Epinal Way} {Loughborough, Leicestershire, LE11 3TU}{S.Ji@lboro.ac.uk} &
  
\end{tabular*}
\addvspace{.75in}

% ---- Two columns begin immediately after authors ----
\begin{multicols*}{2}
\raggedcolumns

% ---- Copyright ----
{\headingfont\bfseries\fontsize{8pt}{12pt}\selectfont
Copyright~\textcopyright~ \the\year{} by the author(s). Permission granted to INCOSE to publish and use.}
\\
% =========================
% ===== Abstract/Keywords =
% =========================
\phantomsection
\miniheading{Abstract}
This paper presents a framework to bridge the gap between subjective stakeholder context and formal system architecture. This is achieved using Soft Systems Methodology (SSM) and Systems Modelling Language version 2 (SysML v2). The methodology utilises the precision of Kernel Modelling Language (KerML) and the alignment of SysML v2 with ISO 42010 to define a reference architecture for the mapping of SSM outputs to SysML v2 concepts such as stakeholders and concerns. Application of the framework is demonstrated through the use of a case study, highlighting the traceable path from stakeholder context to system architecture. The structured mapping and increased semantic precision of SysML v2 are anticipated to reduce the risk of misinterpretation compared to less formal approaches, though empirical validation across diverse stakeholder contexts remains as future work. The primary identified trade-off is the increased barrier to entry associated with SysML v2's textual notation.
% The outputs of the case study show a significantly reduced risk of misinterpretation of the system due to the increased precision of SysML v2, albeit at the risk of increasing the barrier to entry. 

\phantomsection
\subsubsection{Keywords}
MBSE, SysML v2, Soft Systems Methodology, ISO 42010, Stakeholder Context.

% =========================
% ===== Main Content ======
% =========================
\section{1. Introduction}
\label{sec:1}
Soft Systems Methodology (SSM) has long been used by engineers to make sense of complex systems, where objectives may be interdisciplinary, unclear or conflicting (\cite{Checkland1981}). By framing the requirements of a complex system using the viewpoints of the stakeholders involved, SSM creates models that capture system context, which is often missing from hard systems engineering methodologies such as Model-Based Systems Engineering (MBSE). A common and long-standing criticism of MBSE is that, by applying strict logic to ambiguous requirements, the wrong problem is solved effectively (\cite{Mitroff1974}).

While SSM excels at providing context, there remains the need for a formal language to define the resulting system architecture. Historically, the language of choice has been Systems Modelling Language (SysML), an extension of the Unified Modelling Language (UML). However, the software-centric UML was never fully suited to the modelling of physical systems. This disconnect, combined with an ambiguous standard (\cite{Amissah2018}), can lead to an uneven implementation and lack of precision (\cite{Salado2019}). This is mitigated using frameworks such as Unified Architecture Framework (UAF) (\cite{ObjectManagementGroup2022}) to define high-level relationships, although this lacks a procedure for the transition to modelling. Leaving the translation of the high-level stakeholder needs into executable models at the discretion of the individual risks inconsistency between the framework and the implementation (\cite{Morkevicius2021}). 

Prior applications of SSM using SysML have successfully used conceptual models as the basis for both UML (\cite{Bustard1999}) and SysML (\cite{Cloutier2015}) models, providing a starting point to top-level system architecture. However, the suitability of SysML’s semantic structure to capture the conclusions of the conceptual models was questioned, with the enforcement of semantic discipline highlighted as a possible disincentive to using this combination. 

SysML version 2 (SysML v2) addresses the root of this problem by introducing a bespoke metalanguage, Kernel Modelling Language (KerML), as well as a textual notation underpinning a more dynamic version of the graphical representation SysML is known for. SysML v2 is still in the early stages of application and the core semantics of KerML are likely to be refined (\cite{Almeida2025}). Current research includes investigations into the current state of SysML v2’s semantic capabilities (\cite{Litwin2024}) and possible methods of application, such as bespoke profiles (\cite{Kausch2025}), new tools which utilise the textual notation (\cite{Vaicenaviius2025}), and methods of structuring models to optimize data exchange (\cite{Li2024}). A standardised approach for capturing context using SysML v2 concepts remains a significant research gap. Establishing such an approach presents the opportunity to tailor future refinements of KerML and SysML v2 to this specific application.

This paper proposes a framework to capture stakeholder context by providing a reference architecture that maps SSM outputs directly to SysML v2 elements, bridging the gap between ambiguous stakeholder needs and rigorous system architecture. This aligns with the ISO 15288 System Architecture Definition process, providing the transition from stakeholder needs to the requirements definition process (\cite{InternationalOrganizationforStandardization2023}).

The relevant elements of SSM and SysML v2 and their alignment are described, using ISO 42010 (\cite{InternationalOrganizationforStandardization2022}) as an intermediary. This is formalised using a reference architecture (\cite{GitHubAppendices}), the application of which is demonstrated using a generic case study, detailing the transition from stakeholder context to a system architecture in SysML v2.

\section{2. Background}
\label{sec:2}
\subsection{2.1 Soft Systems Methodology}
\label{sec:2.1}
\begin{figure}[H]
  \centering
  \colfig[trim={24cm 2cm 24cm 2cm}, clip]{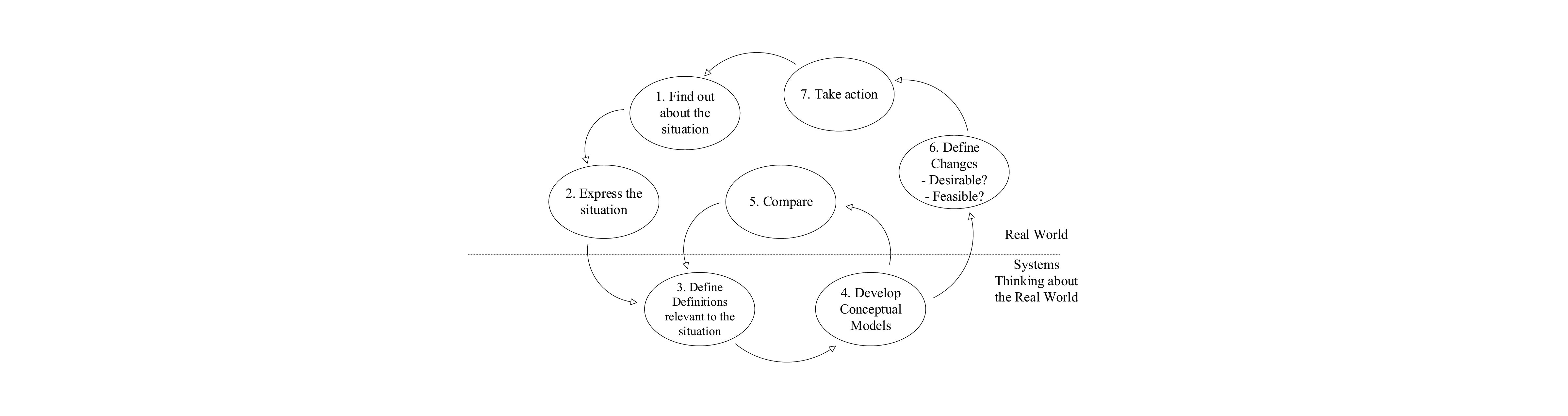}
  \caption{Checkland SSM.}
  \label{fig:Fig1}
\end{figure}

The overall process proposed by Checkland (\cite{Checkland1981}) is shown in Figure \ref{fig:Fig1}.  It is noted that this framework will be familiar to many, as reflected by the age of the citations, but for the sake of context, the areas which are relevant to the proposed framework (1-5) are summarised.

\subsubsection{2.1.1 Rich Picture}
\label{sec:2.1.1}
The goal of the rich picture in SSM is to represent a situation from a number of viewpoints using an illustration, allowing for parallel processing of the information therein (\cite{Wilson2001}) without the constraints of a formal language. This process is flexible; a group of stakeholders may create a rich picture together, or a research team may create one based on interviews (\cite{Mukotekwa2007}), although it has been noted that the latter, while capable of gathering more information, runs the risk of missing key insights (\cite{Gisby2023}). If questions are too open, then key insights may be missed, whilst guiding the direction too forcefully risks introducing the interviewer’s own bias. There is also the risk of an inference gap, where context is misinterpreted or invented by the interviewer, since the stakeholders are not producing the rich picture themselves.

\subsubsection{2.1.2 Root Definitions }
\label{sec:2.1.2}

Wilson describes a Root Definition (RD) as “a way of capturing the essence (root) of the purpose to be served” (\cite{Wilson2001}). A root definition should detail a transformation process, i.e. an input being transformed into an output. In the complex systems that SSM was designed to tackle, there may be vast numbers of these processes, from the high-level to the domain-specific, analytical models (\cite{Checkland1981}). Each of these may then serve a different purpose depending on the viewpoint of the stakeholder who is accessing the system. By defining the individual transformations taking place, the purpose of each process can be defined from the viewpoint of each stakeholder, leading to accurate requirement definition and system structure.

\begin{figure}[H]
  \centering
   \colfig[trim={25cm 1cm 25cm 1cm}, clip]{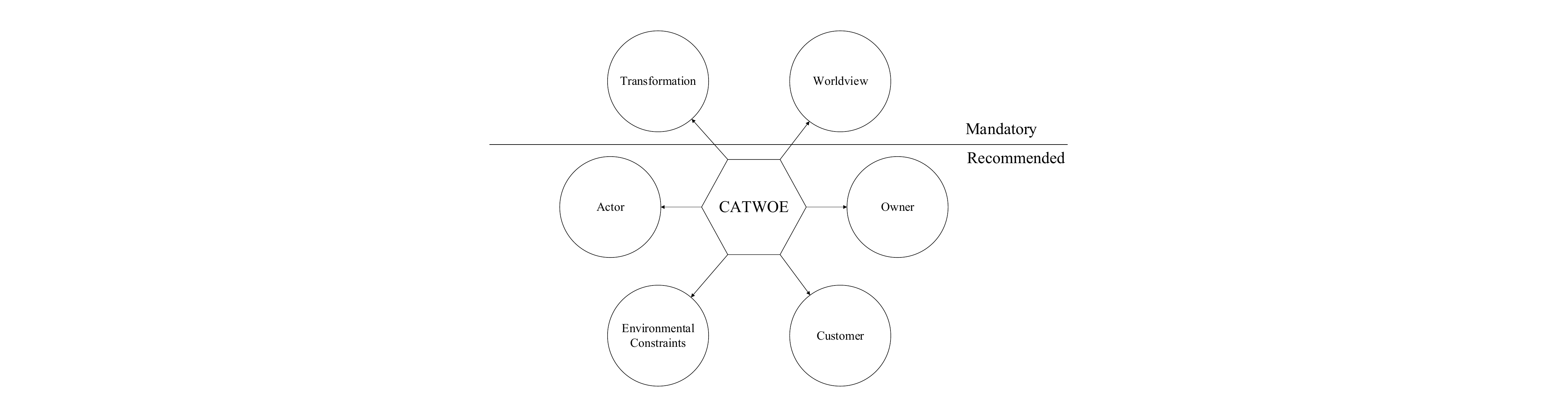}
  \caption{CATWOE Elements.}
  \label{fig:Fig2}
\end{figure}

The application of the CATWOE mnemonic (Figure \ref{fig:Fig2}) serves to define a transformation process from the view of a particular stakeholder. The mandatory elements of the CATWOE are the transformation, defined by a verb, and the worldview, which details what must be believed for the RD to make sense. Individuals are defined as the customer (beneficiary of the transformation), actor (participant in the transformation) and owner (overall decision-maker with performance concerns) elements. By specifying the role an individual plays using the CATWOE structure, their role in the transformation can be defined uniformly. Environmental Constraints are taken as factors significant to the transformation process, which are external to the system itself. 

\subsubsection{2.1.3 Conceptual Models}
\label{sec:2.1.3}

The purpose of the Conceptual Model (CM) is to model the transformation outlined in the RD as a series of activities, including the logical relationships between them, as well as a control sub-system. The purpose of the control sub-system is to define activities which monitor the performance of the activities and act on this to ensure the core activities serve their purpose.

The CM expands on the RD transformation by exploring how to acquire the input, how to reach the output, and how to make the output available for the transformation detailed by the RD (\cite{Wilson2001}). The monitor-and-control system is then applied. These questions are to be answered logically (not based on any pre-existing systems and solutions) to ensure the model remains conceptual. I.e., rather than specifying how or why Person A needs to make money for Person B, the CM would be modelled as a “system to make money”.

The resulting CM should be defensible against the Formal Systems Model (FSM). This ensures that the resulting model describes a “system that is capable of purposeful activity without failure” (\cite{Peters2023}). The application of the FSM is a Verification process (\cite{InternationalOrganizationforStandardization2023}), rather than an Architecture Definition process, falling outside the scope of this framework.

\subsection{2.2 SysML v2}
\label{sec:2.2}

One of the mandatory requirements of SysML v2 was alignment with ISO 42010 (\cite{ObjectManagementGroup2017}). This means the elements and relationships in the SysML v2 specification (\cite{OMG2025}), including stakeholders, concerns and viewpoints, are already positioned to accurately capture the intended meaning of the SSM outputs. To support this, several semantic relationships have been introduced, including:

\begin{itemize}
    \item Definition and Usage

    The Definition and Usage relationship allows for \textit{subsetting}(:>), allowing a usage to be declared as one of the usages of another element, as well as \textit{redefinition }(:>>),  where the inherited feature is replaced by a context-specific usage.
    
    \item Metadata
    
    This replaces the SysML v1 stereotype. Data can now be tagged as inherent properties of an element rather than an annotation.
    
    \item View Specification
    
    The SysML view can now specify filter conditions to query specific parts of the model. This can then be visualised using a specified rendering usage.

    \item Requirements as Constraints

    The SysML v2 requirement is a specialised constraint, which allows for the integration of textual description and logic, creating an executable natural language requirement.
\end{itemize}

As features of SysML v2 are applied during the framework section, context will be added to aid comprehension for newcomers to the language. 

\section{3. Methodology}
\label{sec:3}

The proposed framework seeks to specify a methodology for the uniform modelling of systems using SysML v2, defined first using SSM. This is accomplished in 2 stages: SSM modelling (utilising stakeholder interviews) and the application of a reference architecture for mapping SSM elements to SysML v2 elements. Modelling was completed using CATIA Magic System of Systems Architect, version 2026x.

\subsection{3.1 Rich Picture}
\label{sec:3.1}

\begin{figure}[H]
  \centering
  \colfig[trim={5cm 19cm 5cm 19cm}, clip]{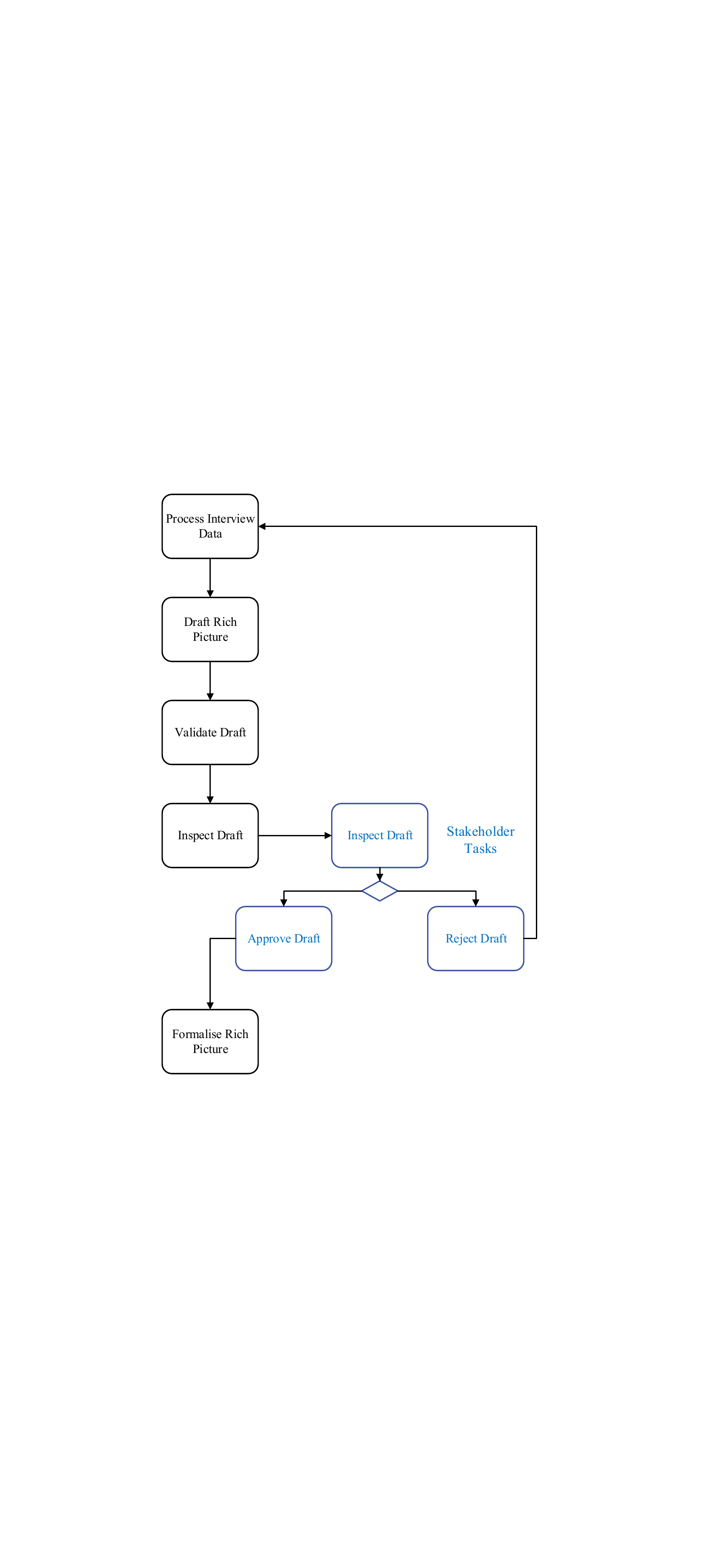}
  \caption{Collaborative Process with Stakeholder Involvement for Rich Picture Development.}
  \label{fig:Fig3}
\end{figure}

When engaging stakeholders to create a rich picture, it is recommended that 1-1 interviews are conducted, with the interviewer then producing a rich picture based on the results. While previously noted that this runs the risk of misinterpretation, there are existing models that can aid the formation of a question set to minimise this risk, ensuring relevant information is acquired. Of these models, this framework recommends the use of POPIT (People, Organisation, Processes, Information, Technology) (\cite{Paul2020}) when forming a question set. Unlike a more traditional People-Process-Tools model (\cite{Leavitt1965}), this ensures that data and artefacts, defined by POPIT as “information”, are considered separately from the technology used to access it. This distinction is crucial, as it enables the  CATWOE Transformation, including inputs and outputs, to be captured accurately.
The rich picture is then created using the interview dataset. The use of targeted questions addresses the risk of lost context during the interviews, but there remains the risk of misinterpretation. It is therefore recommended that the development of the rich picture be an iterative, collaborative process involving the individual stakeholders (Figure \ref{fig:Fig3}).

Creating an initial, individual rich picture for each stakeholder and using follow-up sessions to validate the captured context will improve the accuracy of subsequent modelling and minimise potential rework.  

\subsection{3.2 Root Definitions}
\label{sec:3.2}

\begin{figure}[H]
  \centering
  \colfig{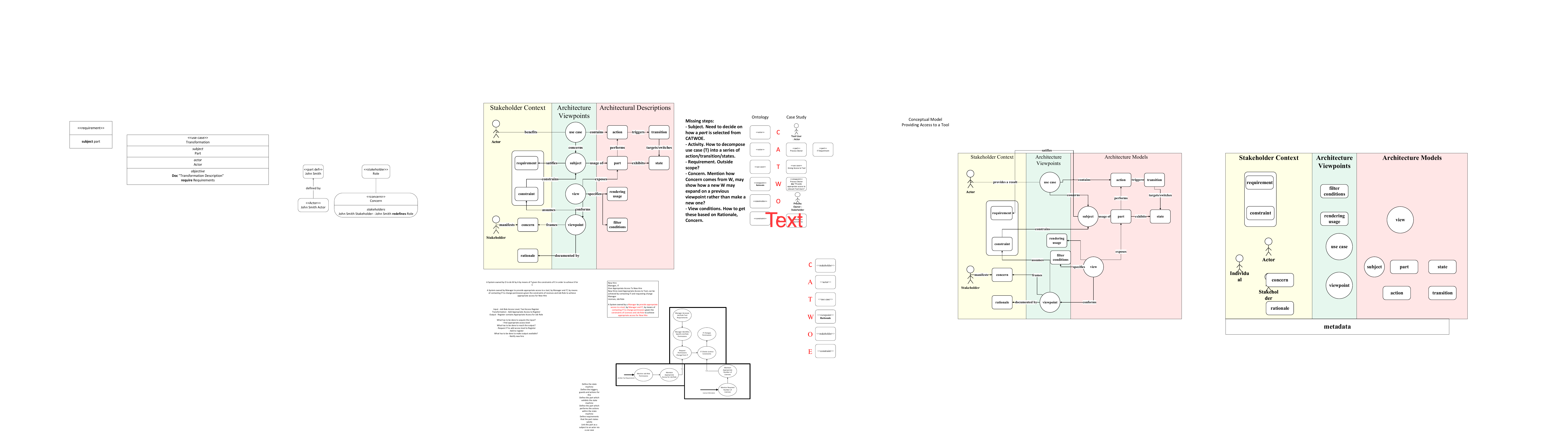}
  \caption{Derivation of key SysML v2 concepts used in the framework.}
  \label{fig:Fig4}
\end{figure}

To ensure the framework for modelling CATWOE elements using SysML v2 is accurate and effective, each CATWOE concept is mapped onto a (or a set of) SysML v2 element. This mapping was based on how the original intent of CATWOE can be represented most accurately using the enhanced semantic discipline of SysML v2, as depicted in the ontological interpretation of a subset of modelling concepts extracted from the SysML v2 metamodel, captured in Figure \ref{fig:Fig4}. 
To take advantage of the alignment with ISO 42010, the relationships contained within the

SysML v2 metamodel has been laid out using the familiar concepts of a stakeholder context domain, architecture models, and the viewpoints which define them. This serves to summarise how different SysML v2 elements relate semantically, as well as defining the overall role they play in the architecture description. 
CATWOE and SysML v2 share common terms, such as \textit{actor}. For clarity, CATWOE elements shall henceforth be capitalised.

A summary of the mapping of SysML v2 and CATWOE elements is provided in Table \ref{tab:Tab1}.

% The [H] option forces the table to stay exactly where you put it in the code.
% Note: This requires \usepackage{float} in your preamble (already in your template).

\begin{table}[H]
    \centering
    \renewcommand{\arraystretch}{1.5} % Makes rows taller for readability
    
    % This uses the custom 'Y' (auto-width) and 'P' (fixed-width) columns defined in your template.
    % If this line causes an error, see the "Troubleshooting" section below.
    \begin{tabularx}{\columnwidth}{@{}P{0.35\columnwidth}Y@{}}
        
        % Header Row
        \rowcolor{tableheader} % Custom gray color from template
        \multicolumn{1}{c}{\headingfont\bfseries CATWOE Element} &
        \multicolumn{1}{c}{\headingfont\bfseries SysML v2 Element} \\ 
        
        % Data Rows
        \addlinespace[4pt] 
        Customer & Individual as Stakeholder. \\
        \addlinespace[4pt]
        Actor & Individual as Actor. \\
        \addlinespace[4pt]
        Transformation & Use Case. \\
        \addlinespace[4pt] 
        Worldview & Viewpoint (with Rationale).\\
        \addlinespace[4pt] 
        Owner & Individual as Stakeholder.\\
        \addlinespace[4pt] 
        Environmental Constraints & Requirements with Constraint.\\
        
    \end{tabularx}
    
    % Caption goes BELOW the table according to your template rules
    \caption{Summary of the SSM - SysML v2 Element mapping.} 
    \label{tab:Tab1}
\end{table}

\subsubsection{3.2.1 Actors}
\label{sec:3.2.1}

%The CATWOE Actors are the individuals who are performing the Transformation. In SysML v1, the actor was a boundary object which lacked internal functional behaviour, which limited traceability from the context to the structure and behaviour of the system. SysML v2 solves this problem by classifying actors as usages of part definitions. As this framework will be modelling real individuals as part definitions, it is useful to include the new individual feature. By specifying a real person as an individual definition, then creating a usage (occurrence) of that definition, SysML v2 allows for the usage to be directly associated with system structure and behaviour (Figure \ref{fig:Fig5}). If this usage is specified within a use case, for example, this usage will only exist within this use case, limiting traceability. To maximise extensibility, the individual should be specified outside the use case. The actor defined within the use case should then redefine (:>>) this high-level usage. This modelling pattern allows for all usages of the individual to be traced back to the occurrence.

\begin{figure}[H]
  \centering
  \colfig{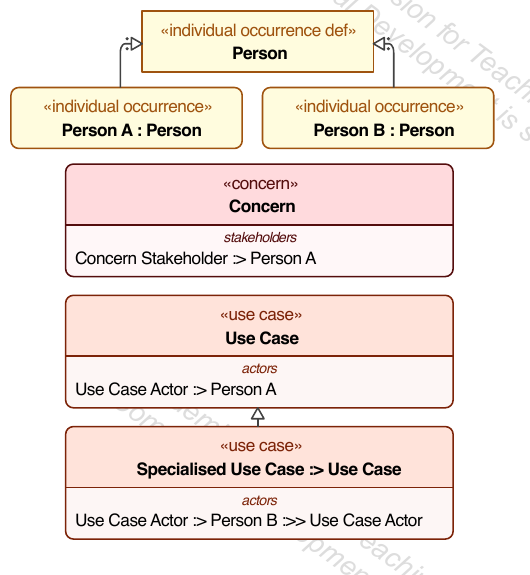}
  \caption{Modelling of an individual both as a stakeholder and an actor.}
  \label{fig:Fig5}
\end{figure}

The CATWOE Actors are the individuals who are performing the Transformation. In SysML v1, the actor was a boundary object which lacked internal functional behaviour, which limited traceability from the context to the structure and behaviour of the system. SysML v2 addresses this problem by classifying actors as usages of part definitions, allowing for local usages to be defined by global definitions. As this framework will be modelling real individuals, it is useful to include the new individual feature. By specifying a person as an individual definition, then creating a usage (occurrence) of that definition, SysML v2 allows for the usage to be directly associated with system structure and behaviour (Figure \ref{fig:Fig5}). If this usage is specified within a use case, for example, this usage will only exist within this use case, limiting traceability. To maximise extensibility, the individual should be specified outside the use case. The local actor usage within the use case should then subset (:>) this high-level usage. This modelling pattern allows for all usages of the individual to be traced back to the occurrence.

\subsubsection{3.2.2 Environmental Constraints}
\label{sec:3.2.2}

In SysML v1, an external constraint could be modelled in one of two ways: as a requirement, which was text-based, or as a constraint block, which contained logic and equations. This meant a trade-off between legibility and functionality. The CATWOE Environmental Constraint is perhaps the element which benefits the most from the switch to SysML v2. Requirements are now defined as specialised constraints, containing both the description of the requirement and at least one constraint. 

SysML v2 constraints can be specified in 3 ways: require, assume and assert. Require constraints serve to formalise the “shall” statement in the requirement. These are generally more suited to system requirements. The assume constraint specifies a list of conditions which must be satisfied before the require constraints apply, useful for modelling assumptions. The assert constraint specifies a condition that must be satisfied at all times. 

The Environmental Constraints should be modelled as requirements, which contain textual information about the constraint. The constraint can then be formalised using mathematics, with the exact application of the constraint decided on a case-by-case basis. By modelling this constraint at the start of the modelling process, it ensures that it can be traced to system elements, and the constraint can be adhered to.

\subsubsection{3.2.3 Transformation and Customer}
\label{sec:3.2.3}

The CATWOE Transformation is defined “either as an input-output conversion or the process itself” (\cite{Wilson2001}). This distinction is important as, following the strict input-output definition, it would suggest the modelling of Transformation as an action. The SysML v2 action is performed by a part, and is defined by its inputs and outputs, which are generally parts or items. The action imposes an effect on its inputs and their parameters to define its outputs. This alone, however, lacks the high-level abstraction captured by the CATWOE Root Definition.

The SysML v2 use case can be used to address the above issue. Specifically, a use case can be used to represent the high-level intent captured by the Transformation, while the actions embedded within the use case can be used to model the actual steps taken to achieve it. These actions are performed by the actors that are associated with the use case via the relationship shown in Figure \ref{fig:Fig5}. The subject of a use case will be the subject of the Transformation, which generally will be a part usage. The use case will also reference a requirement.

As per CATWOE, the beneficiary of the Transformation is the Customer. This will be represented using the SysML v2 stakeholder. A stakeholder has a concern, which is framed by the requirement satisfied by the use case, ensuring complete traceability. The stakeholder embedded in the concern can subset the individual occurrence as in Figure \ref{fig:Fig5}. This allows an individual to act as both a stakeholder and an actor while maintaining traceability to a single source.

\subsubsection{3.2.4 Worldview and Owner}
\label{sec:3.2.4}

In SysML v2, a viewpoint is defined as framing a concern which, according to ISO 42010 (\cite{InternationalOrganizationforStandardization2022}), is a “matter of interest to one or more stakeholders”. Unlike a SysML v2 Actor, which directly participates in the satisfaction of a requirement, a SysML v2 Stakeholder has concerns related to a requirement, to align more closely with ISO 42010. Therefore, the CATWOE Owner can be modelled as a SysML v2 stakeholder, as whilst they have concerns related to the Transformation, they do not (generally) directly perform it as an actor. Maintaining the pattern detailed in Figure \ref{fig:Fig5} can allow an individual to act as both Owner and Actor, but the role of Owner itself is specified using a stakeholder element.

The CATWOE Worldview is a “statement of belief” (\cite{Wilson2001}) stating, from the perspective of the Owner, what the Transformation will achieve and how it will achieve it. As this comes from the point of view of the Owner, this is captured by a viewpoint, but this alone would not fully capture the intent behind the subjective view of the Owner. For example, a viewpoint may state what concern it is addressing, but it lacks the explanation of how and why this viewpoint was chosen.

SysML v2 has introduced rationale as a native element in KerML, as opposed to the somewhat ad-hoc implementation in SysML v1. This allows for textual descriptions to be defined as an inherent property of an element. By including the Worldview as a rationale attached to the viewpoint, the intent of the Owner can be captured, ensuring clear traceability between the implementation and the original intent.

\subsection{3.3 Conceptual Model Mapping}
\label{sec:3.3}

The initial mapping detailed in Table \ref{tab:Tab1} can provide a starting point to begin further system modelling with SysML v2. However, further system structure and behaviour can be modelled using the outputs from the SSM Conceptual Models. This includes the detailed decomposition of the use case, definition of requirements and definition of views.

The theory behind the CM is simple \hyperref[sec:2.1.3]{(2.1.3 Conceptual Models)}. The challenge is to define a method for capturing it completely using SysML v2. It is therefore necessary to elaborate on the key capabilities of the SysML v2 Use Case.

The use case will specify a subject; generally, this is a part. This framework will specify the subject as the part which is being transformed. The exact change in the part can be specified in two ways: using the input and output items of an action to show the change of parameters, or specifying states which the part can exhibit, and modelling the transition between them. The choice of method will be dependent on the Transformation itself, so both methods will be demonstrated in this framework. 
When an action is performed in a use case, an \textit{out} item can be specified. If this item is modelling a signal, it is more useful to model the system separately, using the use case to model the actor’s interaction with the system via that signal. If the item itself is being transformed, then this can be modelled within the use case. Both methods are demonstrated in the following section.

\subsection{3.3.1 Use Case Modelling}
\label{sec:3.3.1}

The SysML v2 part is a kind of item which can perform an action (\cite{OMG2025}). The item element itself can be used to represent the input and output of actions and parts and can have attributes of its own. This naturally aligns with the input and output of the Transformation, so it is recommended that, when specifying the behaviour of a system in a use case, one uses the following structure.

\begin{figure}[H]
  \centering
  \colfig{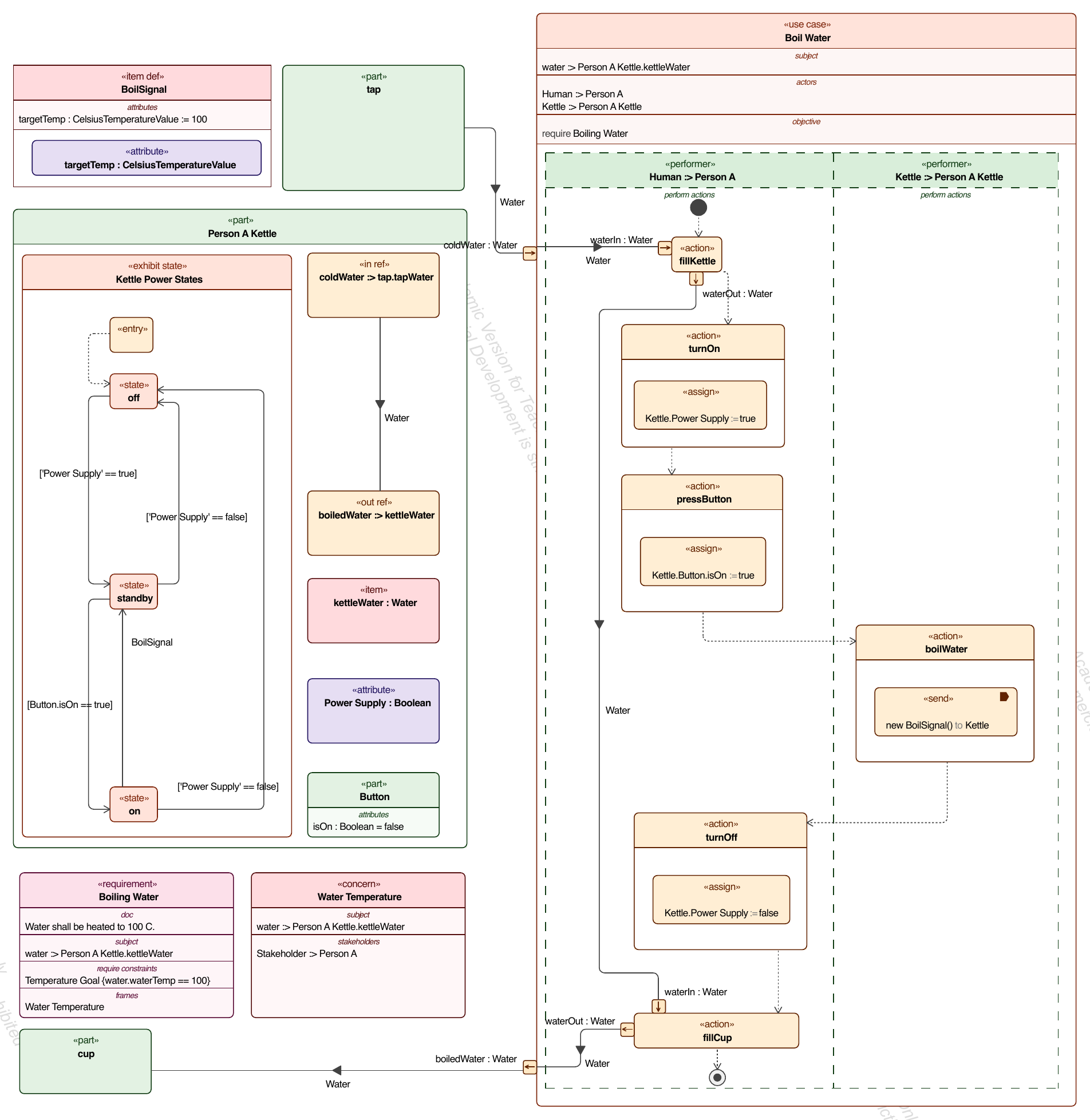}
  \caption{Use Case Modelling.}
  \label{fig:Fig6}
\end{figure}

The key features are detailed in Figure \ref{fig:Fig6}. The simple example of using a kettle to boil water is used. An individual has a concern, which is framed by a requirement. The requirement includes both a description and a mathematical constraint. The subject of the \textit{concern} and the \textit{use case} should be the same to maintain consistency. The \textit{individual} subsets the \textit{actor} usage in the \textit{use case}, performing the \textit{actions} which represent their interaction with the system, with the kettle itself performing the heating \textit{action}. The \textit{use case} references a \textit{requirement} in its \textit{objective}, providing formal traceability from stakeholder concerns to system behaviour.

Two kinds of system behaviour are demonstrated within this \textit{use case}. Actions can have an effect on the \textit{subject} as well as the parts it comprises. The \textit{actions} \textit{turnOn}, \textit{pressButton}, and \textit{turnOff} each contain an \textit{assign}, which modifies \textit{attributes} of the kettle to represent functional changes. Input and output \textit{reference} usages define what enters and exits the system, aligning with the Transformation. 

For more complex systems, it can be more sensible to define structural behaviour separately. A powerful example of this is the SysML v2 state. Figure \ref{fig:Fig6} shows how the state feature, formerly known as the state machine, can be defined within a part to specify its behaviour. Three states are exhibited by the kettle, with transitions defined using guards and by the acceptance of a send action within the use case, both following the standard trigger[guard]/action format. The use case\textit{ }specifies which parts of the system are interacted with, while the behaviour of those parts is specified separately.

%an item, \textit{onSignal}, which is received by an \textit{accept} action within the state \textit{KettleStates}. The use case describes the actor turning the kettle on, so two states are shown to be exhibited by the part: Off and Standby. To change state, a transition is defined in the format \textit{trigger[guard]/action}. In this case, the trigger is \textit{onSignal}, which will turn on the kettle’s standby light, if the power is on. The requirement requires the power to be on and is defined mathematically using a Boolean function. 

\subsubsection{3.3.2 Conceptual Model Mapping}
\label{sec:3.3.2}

Initial modelling details the input and output of the system, i.e., what is being transformed. The Transformation is then elaborated on, to define how the input is acquired, how to reach the output and how to make the output available. This is used to define a series of activities, their ordering and a monitor-and-control subsystem, which is itself made up of activities and relationships. The input and output will be the use case subject, either a part or an item. The activities can be defined as SysML v2 actions, which generate effects on the subject, in the order specified in the CM.

\subsubsection{3.3.3 Requirement Definition}
\label{sec:3.3.3}

The individuals defined by the RD (Actor, Customer and Owner) can all be defined as a stakeholder, actor, or both. The actor is specified in the use case, and the stakeholder is specified in a high-level concern, i.e., safety. A requirement then frames that concern. In practice, however, the requirements will be defined separately. By framing the concern, it then falls under the umbrella of the concern, along with the other requirements which frame safety. 

As previously stated, the CATWOE Environmental Constraints will be modelled using the new requirement-constraint relationship. These will only inform the system from outside the boundary, however. To elicit a complete requirement set, a set of functional requirements should be elicited from the Transformation modelled, as well as the associated elements. Any industry-standard protocol can be followed to complete this and the ensuing non-functional requirements.

\subsubsection{3.3.4 View Definition}
\label{sec:3.3.4}

The primary application of the stakeholder-concern relationship in this framework is the ability to narrow the scope of the total system model to only the elements which are relevant to a particular concern. A SysML v2 view achieves this by exposing a portion of the model, filtering based on metadata, and then rendering in a manner of the modeller’s choosing, such as graphical, tabular, etc. This view then satisfies one or more viewpoints, which in turn frame one or more concerns held by one or more stakeholders.

By specifying the subject of the transformation (use case) as the subject of the concern, traceability from the concern to the model elements is realised, allowing views to be specified according to a stakeholder’s root concern.

\subsubsection{3.3.5 Metadata Definitions}
\label{sec:3.3.5}

SysML v2 introduces metadata definitions, allowing new attribute sets to be defined and used across system elements. This metadata definition allows the specification of attributes of the metadata, such as a textual description. This implementation means that, as opposed to a stereotype or a tag in UML, elements can be queried and filtered within the tool. Usage of metadata in elements will allow a more concrete reference with which filters can render views.

\begin{figure}[H]
  \centering
  \colfig{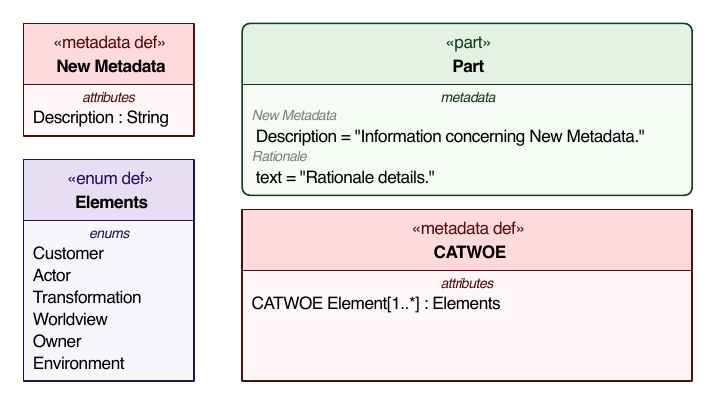}
  \caption{Metadata Modelling.}
  \label{fig:Fig7}
\end{figure}

These definitions can specify attributes, such as a textual description ({Figure \ref{fig:Fig7}}). When specifying a usage of that metadata in an element, that attribute can then be redefined. The definition of Rationale in the SysML v2 specification (\cite{OMG2025}) contains the string attribute “Text” which can be redefined. This is available in the “Modeling Metadata” library package, or can be modelled manually by following the specification. The metadata definition CATWOE contains an attribute “CATWOE Element”, which is typed by a list of enumerations. 

%The \textit{metadata definition }CATWOE is defined with the \textit{attribute }“Element”. \textit{Metadata definitions} for each element were then modelled as \textit{specialisations} of the CATWOE \textit{metadata definition}, with their “Element” \textit{redefined} accordingly. By doing so, model elements representing CATWOE elements can be tagged accordingly, allowing for \textit{filters} to \textit{render views} based on the SSM outputs. 

%Although detailed in the SysML v2 specification, \textit{Rationale} has not yet been implemented at the time of writing this framework. 

%This was overcome by specifying \textit{Rationale }using a \textit{metadata definition}, which included Worldview as an 
%\textit{attribute} (Figure \ref{fig:Fig8}).

\section{ 4. Case Study System}
\label{sec:4}

In the previous section, the technical foundation of the framework was detailed. In order to aid comprehension of the framework, a reference architecture has been populated using the outputs of a case study. This reference architecture and completed case study architecture can be found in the dedicated GitHub Repository (\cite{GitHubAppendices}). Studying the textual notation and graphical views available in the repository is strongly advised to support understanding of the key concepts used in the framework. Descriptions of the foundational concepts are provided in the following sections. SysML v2 elements used for each stage are italicised for differentiation from normal concepts and also for emphasis.

To produce the SSM outputs, a portion of a larger rich picture was taken and anonymised. The overall rich picture was created based on a one-to-one interview with an individual who is a manager in their organisation. The question set used was designed to account for all elements of the POPIT model and is available in the GitHub Repository (\cite{GitHubAppendices}).

\begin{figure}[H]
  \centering
  \colfig[trim={0cm 5cm 0cm 3cm}, clip]{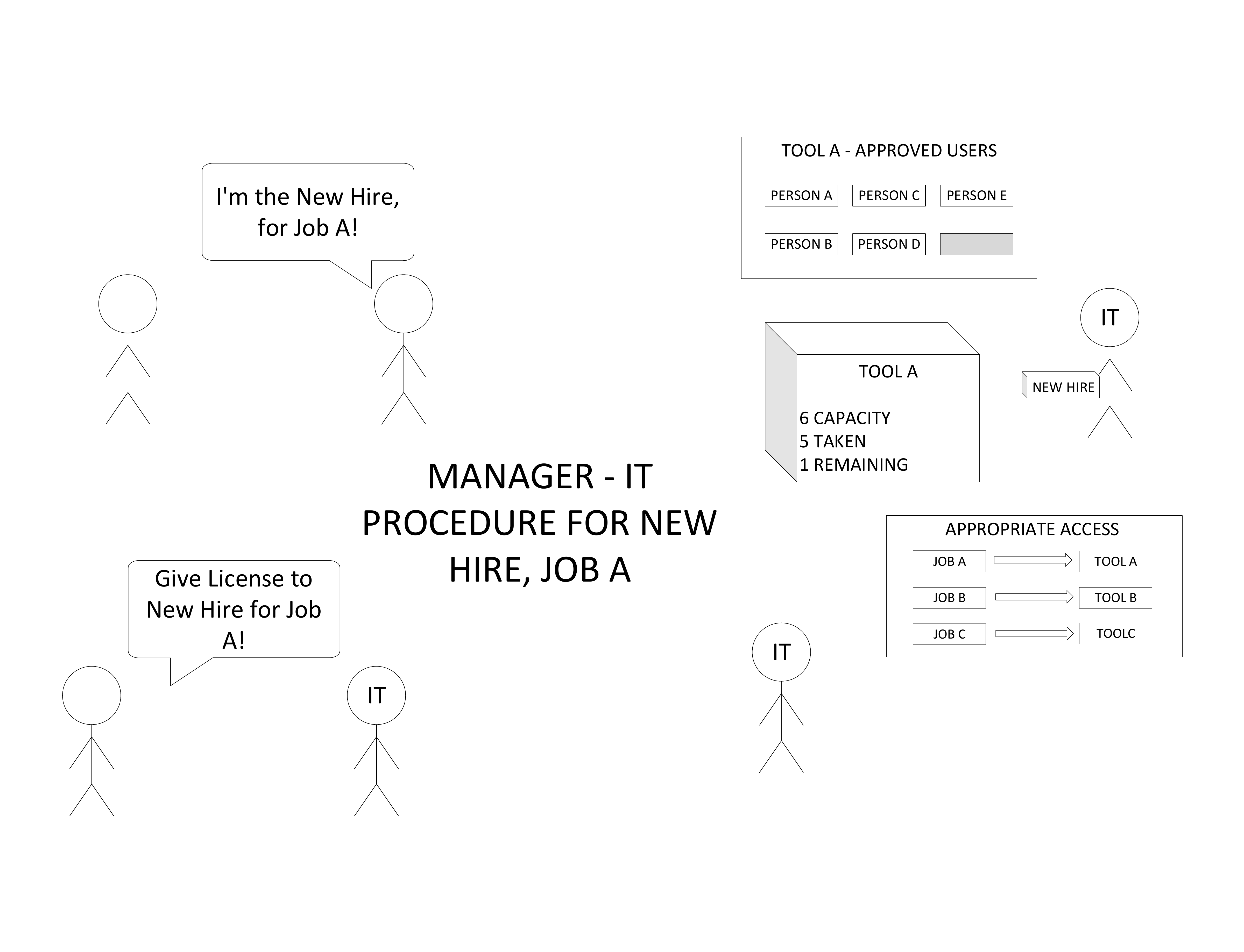}
  \caption{Case Study Rich Picture.}
  \label{fig:Fig8}
\end{figure}

The segment of the rich picture used in this case study (Figure \ref{fig:Fig8}) captured a process which the individual, henceforth referred to as the manager, engages with in their workplace. When a new hire enters the business, they require access to certain tools as per their job responsibilities. The manager then sends a request to the IT department, henceforth referred to as IT, who will assign an available license for the tool that job requires to the new hire. This particular excerpt was chosen as it is a common occurrence in the workplace.

Following the SSM process \hyperref[sec:3.2]{(3.2 Root Definitions)}, the following RD was created:

\begin{narrow}
\textit{A system owned by IT, to maintain appropriate access and license capacity, by IT and the manager, by means of assigning an available license for the correct tool to a new hire, given the constraints of their job role specification and the license availability, to resolve a work request for the manager.}
\end{narrow}

Given that the rich picture intends to capture the system from the worldview of the manager, it can be tempting to assume that the Owner of the process is the manager. However, the rich picture shows that the Transformation, the assignment of a license, is performed by IT, and the action of the manager functions as the trigger of this Transformation. Thus, the Owner of the process should be defined as IT.

Given this RD, the following CATWOE Elements were identified, and then subsequently modelled using SysML v2 based on the mapping defined in \hyperref[sec:3.2]{3.2}:

\begin{itemize}
    \item Customer: Manager
    \item Actors: IT, Manager
    \item Transformation: Assign available license for correct tool to new hire 
    \item Worldview: Appropriate access should be given to new hires and license availability recorded
    \item Owner: IT
    \item Environmental Constraints: Job role specification, license availability  
\end{itemize}

% below has been updated

%\begin{figure*}[t]
 % \centering
 % \colfig{figures/TemplateView.png}
  
 % \caption{Reference Architecture.}
 % \label{fig:Fig19}
%\end{figure*}

\subsection{4.1 Stakeholder Context}
\label{sec:4.1}
\subsubsection{4.1.1 Individuals}
\label{sec:4.1.1}

The three individuals were modelled as \textit{individual occurrences} which were \textit{defined} by the \textit{individual definition} Employee. This \textit{feature typing} relationship allows for the possible extension of traceability and inherited properties at a later stage of model maturity. The \textit{attribute} “name” was modelled at the \textit{definition} level and \textit{redefined} at the \textit{usage} level. The \textit{occurrences} were \textit{tagged} with the appropriate CATWOE role using \textit{metadata}.

\begin{figure}[H]
  \centering
  \colfig{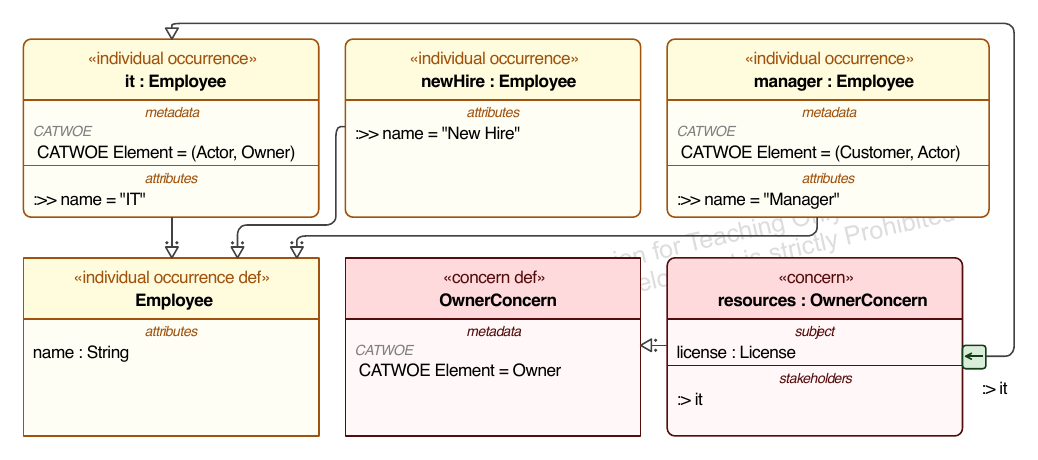}
  \caption{Case Study Individual Modelling.}
  \label{fig:Fig9}
\end{figure}

The central \textit{concern} here was taken as “Resources”, typed by the “OwnerConcern” \textit{definition}. This may be replaced by an existing \textit{concern} if the outputs of this framework are to be integrated into an existing model. IT's role of Owner was then formalised by the local \textit{stakeholder usage} \textit{subsetting} the \textit{occurrence} “it” (Figure \ref{fig:Fig9}).

\subsubsection{4.1.2 System Structure}
\label{sec:4.1.2}

To further elaborate on the stakeholder context, it is useful to begin modelling high-level system elements at this point.

\begin{figure}[H]
  \centering
  \colfig{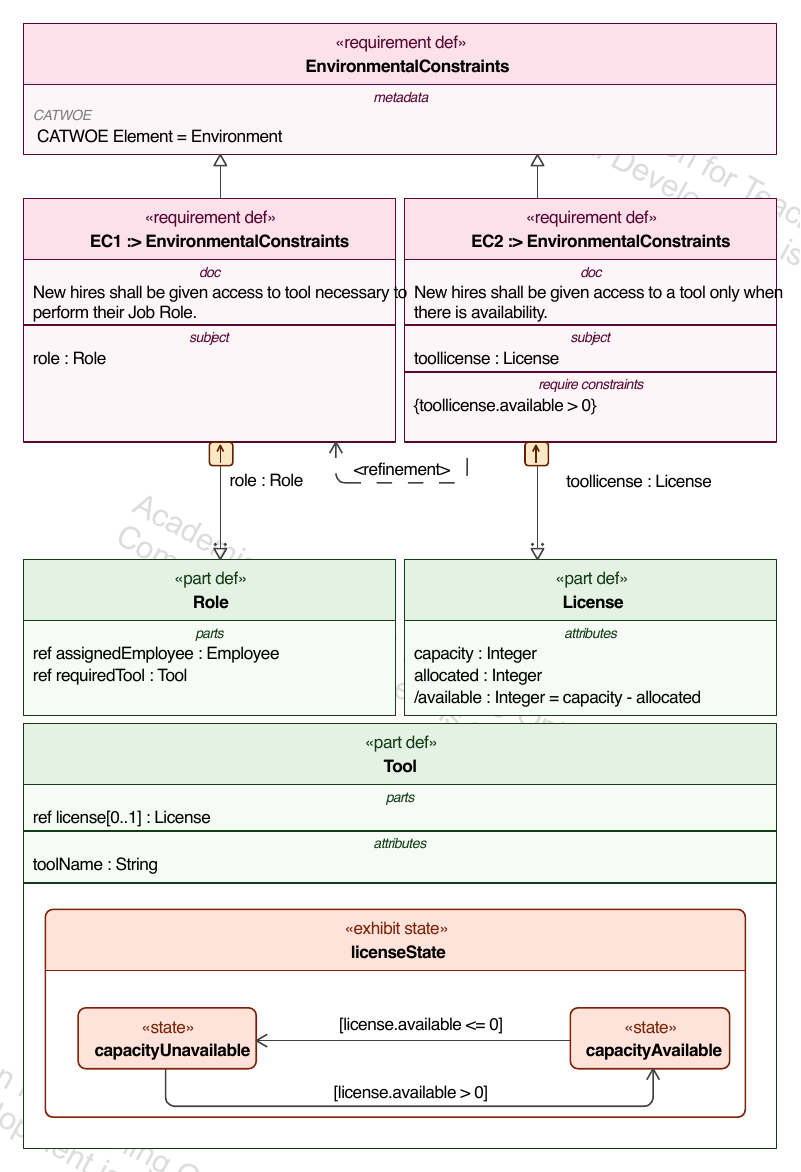}
  \caption{Case Study Initial System Structure - Definition.}
  \label{fig:Fig10}
\end{figure}

\begin{figure}[H]
  \centering
  \colfig{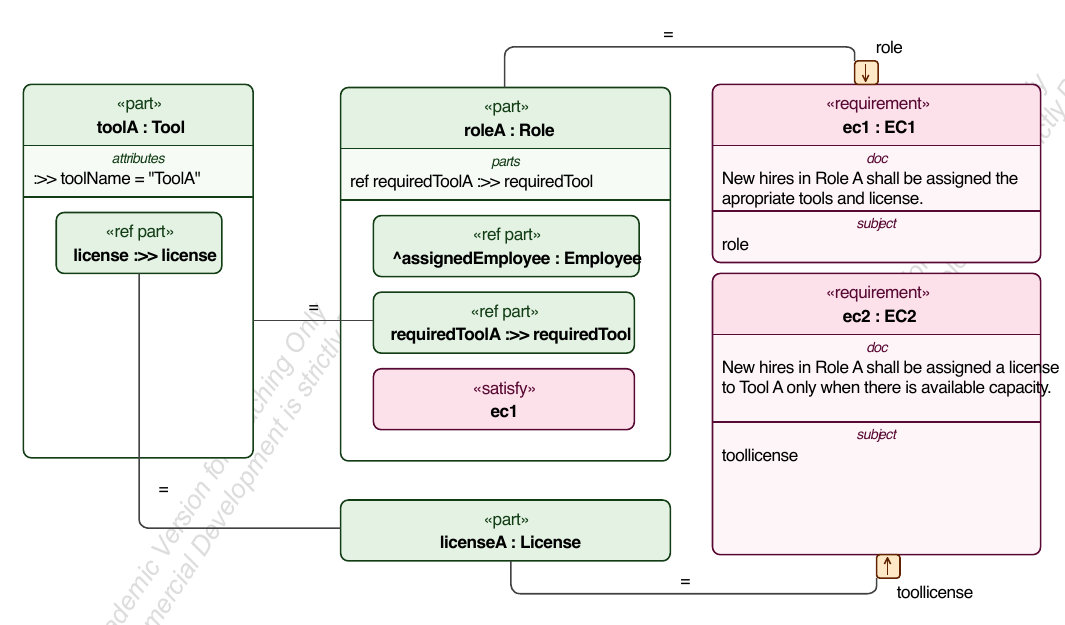}
  \caption{Case Study Initial System Structure - Usage.}
  \label{fig:Fig11}
\end{figure}

From the viewpoint of the manager, the purpose of the system is to give the new hire access to a tool by submitting a request to IT, who will add them to the appropriate tool license according to their job role. This immediately provides the references for three system elements (role, tool and license), which could be modelled using \textit{parts}, as shown in Figure \ref{fig:Fig11}. As these are \textit{part usages}, to ensure model completeness and to enable reuse, a corresponding system element type for each of the \textit{usages} is modelled using \textit{part definition} (Figure \ref{fig:Fig10}). 

%The \textit{feature typing} relationship between the \textit{part definition} and \textit{part usage}s is captured explicitly in Figure \ref{fig:Fig14} to visualise traceability. 

\textit{Reference parts} were modelled at the \textit{definition} stage to explicitly state what type of elements should be referenced. For example, a Tool requires a License, which it may or may not currently have. Therefore, a \textit{reference part} typed by the \textit{part definition} License is modelled, with the optionality captured by the [0..1] multiplicity. These \textit{reference parts} were \textit{redefined} at the \textit{usage} level and associated with the relevant \textit{usage} through \textit{binding}.

It was also noted that when a License is available, there will be a finite capacity. License therefore contains a record of the total capacity and the current allocation, producing a \textit{derived} \textit{attribute} which models the availability. This provides other model elements, such as the \textit{state exhibited} by Tool, with a dynamic level of license availability for use in \textit{guard conditions}.

\subsubsection{4.1.3 Environmental Constraints}
\label{sec:4.1.3}

The RD defines two constraints: the allocated permissions for license access according to the job role and the availability of the associated license. These constraints were modelled as separate \textit{requirement definitions}, typed by \textit{definition} “EnvironmentalConstraints” to inherit the Environment \textit{metadata tag}.

Before modelling the \textit{actions}, the \textit{constraints} of the \textit{use case} must first be specified, as well as the requirements the \textit{use case} will satisfy. 
The first \textit{requirement definition}, EC1, states that “New Hires shall be given access to tools necessary to perform their Job Role”, while the second \textit{requirement definition}, EC2, \textit{refines} the first requirement further, stating that “New hires shall be given access to a tool only when there is availability”. 

%The relationship between the individual, license and tool was previously established by \textit{part usage}, “roleA\_Usage : RoleA”. Therefore, a \textit{satisfy }relationship can be established from the \textit{part} to the \textit{requirement} \textit{usage }of “ec1 : EC1”.

As EC2 concerns the parameter which models license availability, a \textit{require constraint} can be defined (Figure \ref{fig:Fig10}). \textit{Require} was chosen as the \textit{constraint} is the condition to be satisfied by the Transformation, rather than a background assumption.

%This was taken as an \textit{assert} rather than an \textit{assume }(i.e., the Transformation can only take place if the condition is met) because from the view of the stakeholder, this must be true for their Worldview to be satisfied.

\subsection{4.2 Architecture Viewpoints}
\label{sec:4.2}

\subsubsection{4.2.1 Worldview}
\label{sec:4.2.1}

The stakeholder concern, “resources : OwnerConcern”, was \textit{framed} using a \textit{viewpoint}, “licenseManagement : ResourceAllocation”. This captures the Worldview using the \textit{Rationale} \textit{metadata}. The defining \textit{viewpoint definition} was \textit{tagged} with the “Worldview” \textit{metadata}, allowing for this to be captured in the \textit{usage}.

A \textit{view}, “License Allocation”, was then modelled, which \textit{satisfies viewpoint} “licenseManagement : ResourceAllocation”. This view can \textit{expose} specific system \textit{elements} and \textit{filter} them based on \textit{metadata} and \textit{typing}. This was not found to be sufficiently mature enough to define a generic \textit{view} specification, so was left blank as an illustrative \textit{view}.

\begin{figure}[H]
  \centering
  \colfig{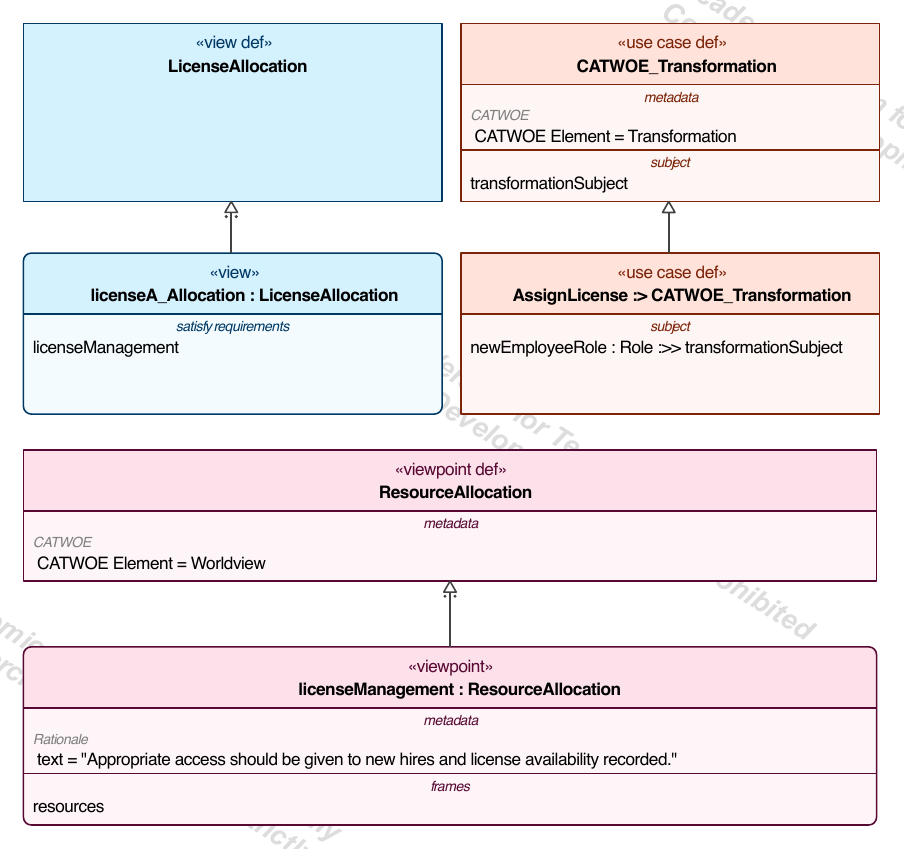}
  \caption{Architecture Viewpoint Modelling.}
  \label{fig:Fig12}
\end{figure}

\subsubsection{4.2.2 Transformation}
\label{sec:4.2.2}

The high-level Transformation was defined using a \textit{use case}, with the local \textit{actor usages} \textit{subsetting} \textit{individuals} “it” and “manager”. This \textit{use case} was \textit{typed} by \textit{use case definition} “AssignLicense”, which specifies that the \textit{subject} must be typed by the \textit{definition} “Role”. “AssignLicense” is typed by the \textit{use case definition} “CATWOE\_Transformation” and \textit{tagged} using the Transformation CATWOE label.

\begin{figure}[H]
  \centering
  \colfig{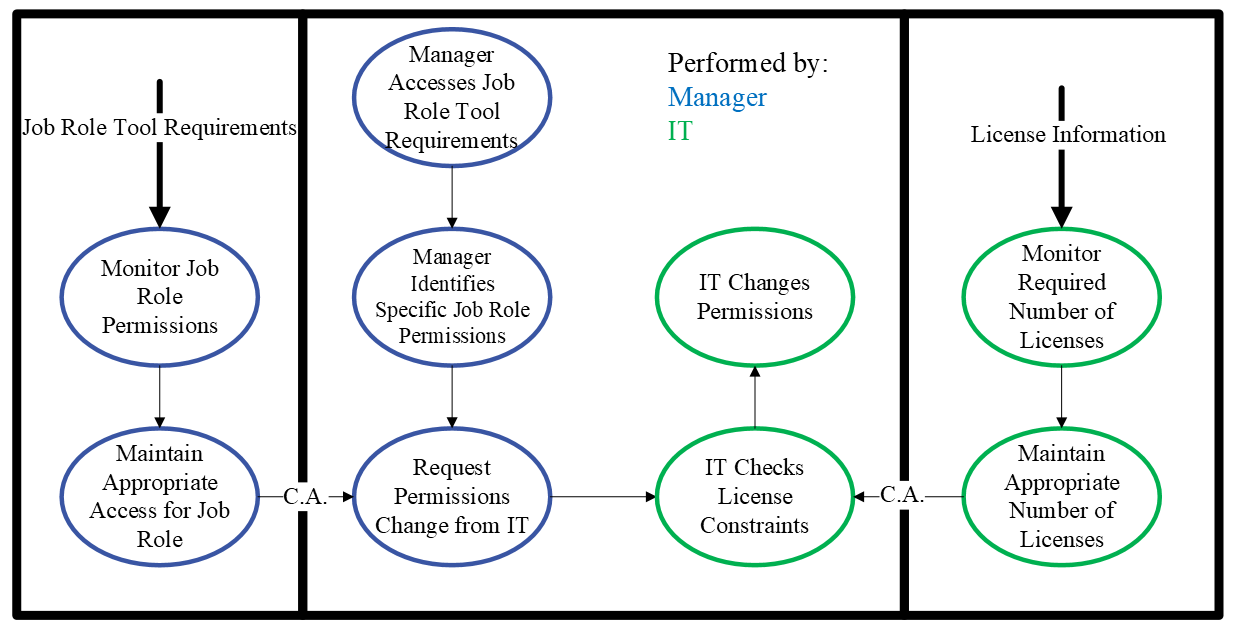}
  \caption{Case Study Conceptual Model.}
  \label{fig:Fig13}
\end{figure}

At this stage, all six CATWOE elements have been modelled and \textit{tagged} using \textit{metadata}. To further decompose the Transformation and system structure, the CM can be referred to (Figure \ref{fig:Fig13}), which was 

developed using the process described previously \hyperref[sec:2.1.3]{(2.1.3 Conceptual Models)}. The central activity system details the activities which comprise the Transformation and were assigned to either the manager or IT by colour. These activities were modelled using \textit{actions}. Two monitor-and-control systems are also included, which specify necessary control actions, as well as required sources of information. These may inform the further elicitation of requirements and their formalised constraints, although this was not included in this stage of the modelling process.

According to the RD, it is known that IT will add the new hire to the appropriate license, but the specifics are not known. Therefore, the implementation process is left abstract during modelling, with greater focus on the information which will be required to enable each of the activities. If, during specific applications of this framework, a greater level of detail is required, it is recommended that this is developed in collaboration with the relevant stakeholder as per Figure \ref{fig:Fig3}. This will ensure that the resulting solution meets the specific needs of the stakeholder, eliminating the possibility of wasted effort and rework.

\begin{figure*}[t]
  \centering
  \colfig{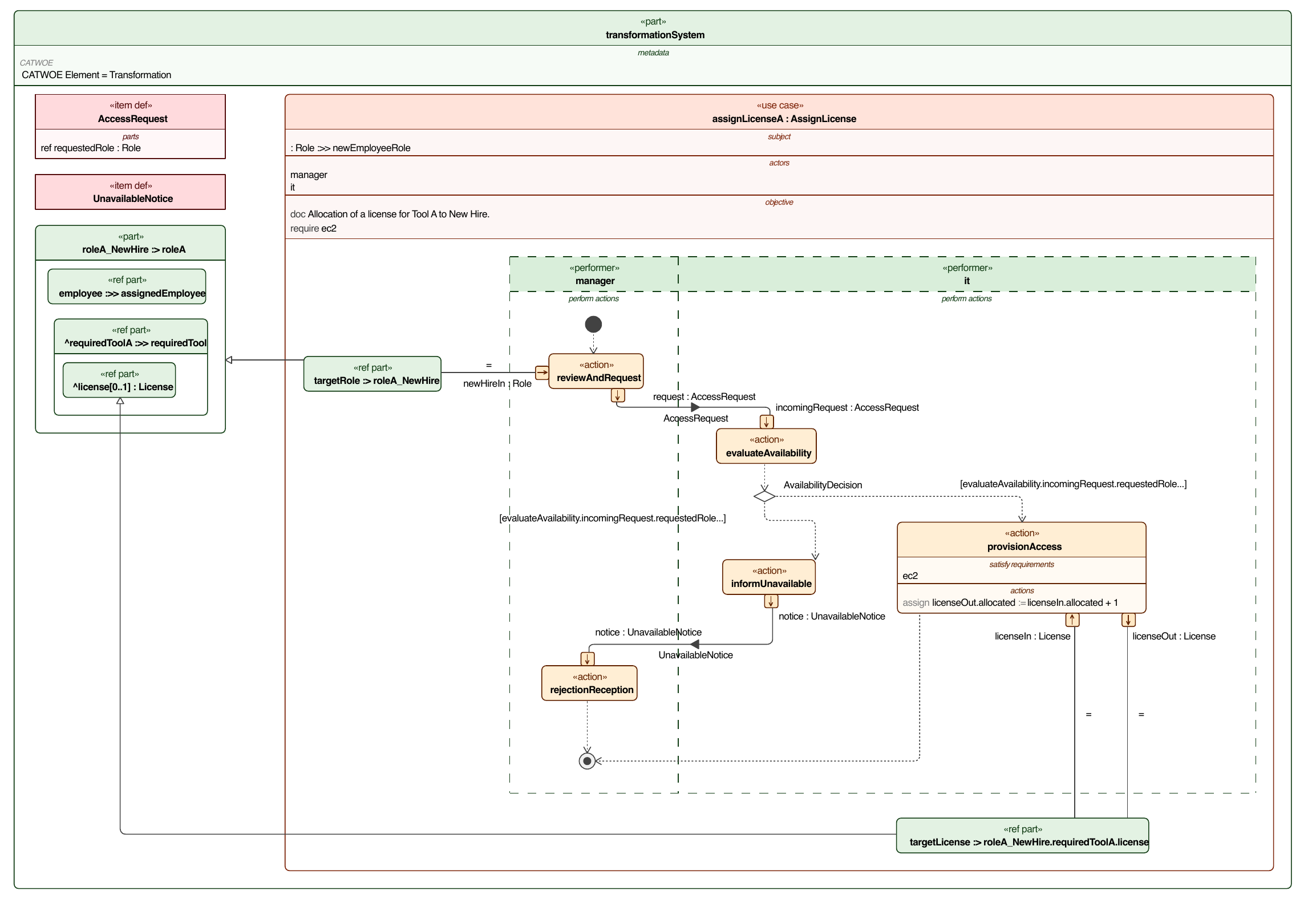}
  
  \caption{Case Study Transformation Use Case.}
  \label{fig:Fig14}
\end{figure*}

\subsection{4.3 Architectural Descriptions}
\label{sec:4.3}
\subsubsection{4.3.1 Use Case}
\label{sec:4.3.1}

To model the \textit{usage} of this \textit{use case}, a \textit{subject} must be selected. A part “transformationSystem” was created, which includes the \textit{use case usage} and the “Role A” usage to be occupied by New Hire, “roleA\_NewHire”, which fills the role of \textit{subject} via \textit{subsetting}.  The overall \textit{part} “transformationSystem” details the process by which the Transformation is performed, including the sources of information required by each \textit{action}.  By modelling a \textit{part} which includes the \textit{use case} as the Transformation, rather than the \textit{use case} alone, the full Transformation process and all necessary elements can be captured, answering the three questions previously described \hyperref[sec:2.1.3]{(2.1.3)}. This top-level \textit{part} can also be exposed to generate the \textit{view}, although the level of specificity shown in Figure \ref{fig:Fig14} currently cannot be achieved through the generic use of \textit{filter} and \textit{expose} alone.

%figure updated to increase line thickness

The \textit{use case} details illustrative steps for completing the Transformation, with \textit{actions} being \textit{performed} by \textit{actors}. The exact location of different attributes is also defined using \textit{binding} to \textit{reference usages}. The two branches of the \textit{decision} result in different outputs, one with a \textit{message}, the other with an \textit{assign action} which alters an \textit{attribute}. By explicitly linking the possible scenarios to model elements, the accuracy and comprehensibility of the model are increased.

%The \textit{action} “checkLicenseConstraint” contains a \textit{guard} to only proceed if the \textit{constraint} in "ec2 : EC2" is satisfied. The \textit{action} “changePermissions” may specify a script to change the “allocated” \textit{attribute} of “licenseA”, although this could not be completed due to tool limitations at the time of writing. The product of the \textit{use case} is the \textit{redefined} (:>>) \textit{part usage} of “RoleA”, now with allocated employee “newHire”. 

\section{5. Conclusions}
\label{sec:5}

The application of the framework to a case study revealed the following findings. 

Firstly, the textual notation has greatly enhanced the ease of precise modelling, making the use of feature typing and redefinition a natural part of the modelling process. Beyond using the textual notation itself, the language also allows the modeller to verify traceability when modelling graphically. As the size of the textual notation grew, however, it became more important to rely on the graphical representation to keep a clearer image of the entire model. This may present a barrier to entry for less experienced modellers, although it is anticipated that developments in the use of Large Language Models (LLMs) will address this as the area of research matures. The use of a framework provides a structured mapping of concepts, creating an ideal environment for an LLM to be leveraged. This could include automated generation of new model elements which adhere to the framework, natural language explanations of the modelled concepts, and facilitating the definition of complex views. The graphical representation was found to be flexible, allowing for the modeller to present the system in the manner of their choosing without interacting with the underlying logic. However, it is noted that this risks misrepresenting the system structure and behaviour. This demonstrates further the requirement for the formalisation of view definitions using SysML v2. By applying filtering conditions, the benefits of the graphical notation can be maintained whilst helping an individual modeller avoid misrepresenting an already-defined system.

The use of the SSM process was decidedly useful in producing a formalised and reproducible set of initial system elements. Using a reference architecture to model this was also successful. By providing the outputs, as well as the template and SSM outputs, to a subject matter expert or external entity responsible for modelling, the likelihood of \textit{accurate} models should be increased.

%\begin{figure*}[t]
 % \centering
  %\colfig[trim={1.2cm 54cm 1cm 1cm}, clip]{figures/Fig14.pdf}
  
  %\caption{Case Study System Architecture.}
  %\label{fig:Fig14}
%\end{figure*}

%updated below

%\begin{figure*}[t]
 % \centering
  %\colfig{figures/CaseStudy.png}
  
  %\caption{Case Study System Architecture.}
  %\label{fig:Fig14}
%\end{figure*}

There were two significant challenges that were left unaddressed, however: further modelling of system elements, and integration with existing system elements. The use of a standardised format and a template should assist with the integration with other systems using the framework, but the integration with systems modelled using another framework, or no framework, should be investigated. 

This study has established a framework for applying Soft Systems Methodology practices and formalising its outputs using SysML v2 in a uniform and reasoned manner. By mapping CATWOE elements to specific SysML v2 elements via the reference architecture and applying this to a case study, the risk of solving the wrong problem is addressed, as the context is no longer ambiguous. This allows accurate communication of that context and, therefore, more accurate MBSE models. Whilst it was noted that the balance between textual and graphical modelling may require some nuance and expertise, it ultimately made enforcing precision on the model a more inherent part of the process, which can be carried into subsequent modelling activities.

\section{6. Future Directions}
\label{sec:6}

This framework has set out a methodology for formalising stakeholder context within the current limitations of initial SysML v2 deployment. To further develop the capabilities of the framework, it is necessary to explore the filter conditions of the view element, as well as how to query the model. By exploring these capabilities, extended frameworks can be defined for specific modelling applications, ensuring the accuracy of the graphical notation. 

Validation against the Formal Systems Model (FSM) was not completed as part of this framework. However, it is recommended that the SSM protocol include an FSM comparison in future extensions of this framework. The resulting use case modelling process can then be compared to that of this framework to explore potential benefits of an expanded SSM protocol, potentially offering increased levels of resolution. 

Thirdly, it is recommended that a case study be conducted investigating the application of this framework by individuals of varying MBSE and SSM experience. The benefits previously outlined may differ in weight depending on the modelling skill of the user. By doing so, the results can be compared and act as a verification and validation process for the framework.

Lastly, the monitor-and-control loops in Figure \ref{fig:Fig13} have not been modelled in detail, to avoid confusion for the reader. It is recommended that the framework be extended to include a separate procedure for this using a sophisticated monitor-and-control system. This will further explore the capabilities of SysML v2 whilst also providing a second model, which can be used to investigate the integration of two systems modelled using this framework.

\section{Acknowledgments}

Gemini was used for initial checking of grammar and continuity of figures. Corrections were verified before being applied by the authors.

\printbibliography[]

@misc{InternationalOrganizationforStandardization2023,
   author = {{International Organization for Standardization} and {International Electrotechnical Commission} and {Institute of Electrical and Electronics Engineers}},
   city = {Piscataway, NJ, USA},
   doi = {10.1109/IEEESTD.2023.10123367},
   isbn = {978-1-5044-9665-0},
   month = {3},
   publisher = {IEEE},
   title = {{ISO}/{IEC}/{IEEE} International Standard - Systems and software engineering--System life cycle processes},
   year = {2023}
}

@article{Morkevicius2021,
   author = {Aurelijus Morkevicius and Aiste Aleksandraviciene and Gintare Krisciuniene},
   doi = {10.1002/j.2334-5837.2021.00856.x},
   issn = {2334-5837},
   number = {1},
   journal = {INCOSE International Symposium},
   title = {From {UAF} to {SysML}: Transitioning from System of Systems to Systems Architecture},
   volume = {31},
   year = {2021}
}

@article{Salado2019,
   author = {Alejandro Salado and Paul Wach},
   doi = {10.3390/systems7020019},
   issn = {20798954},
   number = {2},
   journal = {Systems},
   title = {Constructing true model-based requirements in {SysML}},
   volume = {7},
   year = {2019}
}

@article{Amissah2018,
   author = {Matthew Amissah and Ange-Lionel Toba and Holly A H Handley and Mamadou Seck},
   journal = {SpringSim},
   title = {TOWARDS A FRAMEWORK FOR EXECUTABLE SYSTEMS MODELING: AN EXECUTABLE SYSTEMS MODELING LANGUAGE ({ESYSML})},
   year = {2018}
}

@article{Li2024,
   author = {Zirui Li and Faizan Faheem and Stephan Husung},
   doi = {10.3390/systems12010018},
   issn = {20798954},
   number = {1},
   journal = {Systems},
   title = {Collaborative Model-Based Systems Engineering Using Dataspaces and {SysML} v2},
   volume = {12},
   year = {2024}
}

@article{Vaicenaviius2025,
   author = {Juozas Vaicenavičius and Tilo Wiklund and Daumantas Kavolis and Simonas Draukšas and Antanas Kalkauskas and Rimantas Vaicenavičius},
   doi = {10.1007/s12567-025-00595-x},
   issn = {18682510},
   journal = {CEAS Space Journal},
   title = {{SysIDE}: {SysML} v2 textual editing and analysis system: overview and applications},
   year = {2025}
}

@article{Kausch2025,
   author = {Hendrik Kausch and Mathias Pfeiffer and Deni Raco and Bernhard Rumpe and Andreas Schweiger},
   doi = {10.1007/s13272-024-00762-6},
   issn = {18695590},
   number = {1},
   journal = {CEAS Aeronautical Journal},
   title = {Model-driven development for functional correctness of avionics systems: a verification framework for {SysML} specifications},
   volume = {16},
   year = {2025}
}

@inproceedings{Litwin2024,
   author = {Kyle Litwin and Isaac Amundson and Dinesh Verma and Tom McDermott},
   doi = {10.4271/2024-01-1947},
   issn = {01487191},
   booktitle = {SAE Technical Papers},
   title = {Transforming {AADL} Models into {SysML} 2.0: Insights and Recommendations},
   year = {2024}
}

@inproceedings{Almeida2025,
   author = {João Paulo A. Almeida and Luís Ferreira Pires and Giancarlo Guizzardi and Gerd Wagner},
   doi = {10.1007/978-3-031-75872-0_8},
   issn = {16113349},
   booktitle = {Lecture Notes in Computer Science (including subseries Lecture Notes in Artificial Intelligence and Lecture Notes in Bioinformatics) },
   title = {An Analysis of the Semantic Foundation of {KerML} and {SysML} v2},
   volume = {15238 LNCS},
   year = {2025}
}

@inproceedings{Bustard1999,
   author = {D. W. Bustard and Z. He and F. G. Wilkie},
   doi = {10.1109/hicss.1999.772894},
   issn = {10603425},
   booktitle = {Proceedings of the Hawaii International Conference on System Sciences},
   title = {Soft systems and use-case modelling: Mutually supportive or mutually exclusive?},
   year = {1999}
}

@techReport{ObjectManagementGroup2022,
   author = {{Object Management Group}},
   institution = {Object Management Group},
   title = {Unified Architecture Framework ({UAF}) Domain Metamodel ({DMM}) Version 1.2},
   url = {https://www.omg.org/spec/UAF/1.2},
   year = {2022}
}

@book{InternationalOrganizationforStandardization2022,
   author = {{International Organization for Standardization} and {International Electrotechnical Commission} and {Institute of Electrical and Electronics Engineers}},
   publisher = {ISO/IEC/IEEE},
   title = {Software, systems and enterprise — Architecture description},
   url = {https://www.iso.org/standard/74393.html},
   year = {2022}
}

@techReport{ObjectManagementGroup2017,
   author = {{Object Management Group}},
   institution = {Object Management Group},
   title = {Object Management Group Systems Modeling Language ({SysML}®) v2 Request For Proposal ({RFP})},
   url = {http://www.omg.org/cgi-bin/doc.cgi?ad/2017-12-2},
   year = {2017}
}

@inbook{Leavitt1965,
   author = {Harold J. Leavitt},
   city = {Chicago},
   editor = {James G. March},
   booktitle = {Handbook of Organisations},
   pages = {1144-1170},
   publisher = {Rand McNally},
   title = {Applied Organizational Change in Industry: Structural, Technological and Humanistic Approaches},
   year = {1965}
}

@book{Paul2020,
   author = {Debra Paul and James Cadle and Malcolm Eva and Craig Rollason and Jonathan Hunsley},
   edition = {4},
   publisher = {BCS, The Chartered Institute for IT},
   title = {Business Analysis},
   year = {2020}
}

@book{Wilson2001,
   author = {Brian Wilson},
   edition = {1},
   publisher = {John Wiley \& Sons},
   title = {Soft Systems Methodology: Conceptual Model Building and Its Contribution},
   year = {2001}
}

@misc{OMG2025,
   author = {{Object Management Group}},
   title = {About the {OMG} System Modeling Language Specification Version 2.0},
   url = {https://www.omg.org/spec/SysML/2.0/},
   year = {2025}
}

@book{Checkland1981,
   author = {Peter Checkland},
   publisher = {John Wiley \& Sons},
   title = {Systems Thinking, Systems Practice},
   year = {1981}
}

@article{Mitroff1974,
   author = {Ian I. Mitroff and Tom R. Featheringham},
   doi = {10.1002/bs.3830190605},
   issn = {10991743},
   number = {6},
   journal = {Behavioral Science},
   title = {On systemic problem solving and the error of the third kind},
   volume = {19},
   year = {1974}
}

@article{Peters2023,
   author = {Geoff Peters and Joyce Fortune and Diana White},
   doi = {10.54120/jost.0000010},
   number = {1},
   journal = {Journal of Systems Thinking},
   title = {The Formal System Model},
   volume = {3},
   year = {2023}
}

@article{Mukotekwa2007,
   author = {Charity Mukotekwa and Ewart Carson},
   doi = {10.1177/1744987107078897},
   issn = {17449871},
   number = {6},
   journal = {Journal of Research in Nursing},
   title = {Improving the discharge planning process: A systems study},
   volume = {12},
   year = {2007}
}

@article{Gisby2023,
   author = {Alison Gisby and Catharine Ross and Jan Francis-Smythe and Kazia Anderson},
   doi = {10.1177/15344843221148044},
   issn = {15526712},
   number = {2},
   journal = {Human Resource Development Review},
   title = {The ‘Rich Pictures’ Method: Its Use and Value, and the Implications for {HRD} Research and Practice},
   volume = {22},
   year = {2023}
}

@article{Cloutier2015,
   author = {Robert Cloutier and Brian Sauser and Mary Bone and Andrew Taylor},
   doi = {10.1109/TSMC.2014.2379657},
   issn = {21682232},
   number = {4},
   journal = {IEEE Transactions on Systems, Man, and Cybernetics: Systems},
   title = {Transitioning systems thinking to model-based systems engineering: Systemigrams to {SysML} models},
   volume = {45},
   year = {2015}
}

@software{GitHubAppendices,

  title = {Github Repository},
author = {Harrison, Matthew},
  year = {2025},
  url = {https://github.com/maatt1199/SSM-SysML-v2-Reference-Architecture},
  version = {1.0}
}

\end{multicols*}

\end{document}